\documentclass[twocolumn,showpacs,preprintnumbers,amsmath,amssymb,aps]{revtex4}
\usepackage{graphicx}

\begin{document}

\title{Reversible Vortex Ratchet Effects and Ordering   
in Superconductors with 
%One-Dimensional 
Simple Asymmetric Potential Arrays}   
\author{Qiming Lu$^{1,2}$, C.J. Olson Reichhardt$^{1}$, and 
C. Reichhardt$^{1}$ }
\affiliation{ 
{$^1$}Center for Nonlinear Studies 
and Theoretical Division, 
Los Alamos National Laboratory, Los Alamos, New Mexico 87545\\
{$^2$}Department of Physics, Applied Physics, \& Astronomy, Rensselaer 
Polytechnic Institute, Troy, New York, 12180-3590} 

\date{\today}
\begin{abstract}
We demonstrate using computer simulations
that the simplest vortex ratchet system 
for type-II superconductors with artificial pinning arrays, a
%n 
simple asymmetric
%one-dimensional (1D) 
potential array, exhibits the same features as more 
complicated
two-dimensional (2D) vortex ratchets that have been studied in recent 
experiments.
We show that the 
%1D 
simple geometry, originally proposed by Lee {\it et al.} 
[Nature {\bf 400}, 337 (1999)], 
undergoes multiple reversals in the sign of the 
ratchet effect as a function of
vortex density, substrate strength, and ac drive amplitude, and that 
the sign of the ratchet effect is related to the type of vortex lattice
structure present. Thus, although the ratchet geometry 
has the appearance of being effectively 1D, the behavior
of the ratchet is affected by the 2D structure of the vortex configuration.
When the vortex lattice is highly ordered, an 
ordinary vortex ratchet effect occurs which is similar to the 
response of an isolated particle in the same ratchet geometry.  In regimes
where the vortices form a smectic or disordered phase, 
the vortex-vortex interactions are relevant 
and we show with force balance arguments that 
the ratchet effect can reverse in sign.   
The dc response of this system features a reversible diode effect   
and a variety of vortex states including 
triangular, smectic, disordered and square.  
\end{abstract}
\pacs{74.25.Qt}
\maketitle

\vskip2pc 

\section{Introduction} 
When an overdamped particle is placed in an asymmetric 
potential and an additional ac drive is applied, a net 
dc drift velocity or rectification can occur 
which is known as the ratchet effect.
Stochastic ratchets can be constructed with Brownian particles, while
ordinary ratchets can be created in deterministic systems
\cite{Reimann}.  
Typically,
an applied ac drive or periodic flashing of the
potential couples with some form of asymmetry in the substrate, 
breaking the symmetry of the particle motion.
Ratchet effects have been      
studied in the context of molecular motors \cite{Reimann} 
as well as in the motion of colloidal particles \cite{Colloids},
granular materials \cite{Farkas} and 
cold atoms in optical trap arrays \cite{Wallin}. 

Vortices in type-II superconductors act as overdamped
particles moving in nanostructured 
potential landscapes.
The possibility of realizing a ratchet effect to transport 
superconducting vortices was  
originally proposed by Lee {\it et al.} \cite{Janko}, who studied vortices  
interacting with a 
%1D 
simple asymmetric or sawtooth substrate potential 
and an additional external ac drive. 
This type of asymmetric potential could be fabricated by creating
a superconducting sample that has an asymmetric thickness modulation.
The external ac drive is provided by either an oscillating external 
magnetic field or by an
applied ac current which produces an oscillating Lorentz 
force on the vortices.
The numerical simulations of Ref.~\cite{Janko} indicated
that a net dc motion of vortices could arise upon application of
an ac drive, and that the net vortex drift is in the
easy direction of the asymmetric substrate, 
as would be expected for a single particle interacting with such a
potential. 
The original work on vortex ratchets was followed by
a series of proposals for alternative vortex ratchets based on
2D asymmetric pinning array geometries, each 
of which produces similar types of vortex ratchet effects with
the net vortex flow 
in the easy direction of the underlying potential 
\cite{Wambaugh,Olson,Morelle,Togawa,Wu}.
%NEW
These ratchets operate either through a bottleneck effect 
\cite{Wambaugh,Togawa} or by means of an effectively 1D mechanism
\cite{Olson,Morelle,Wu}.
%END NEW
Similar ratchet effects have also been found for superconductors with 
asymmetric surface barriers \cite{Peeters}. 

Recently, new types of
2D periodic pinning geometries have been proposed that 
exhibit not only the regular ratchet effect but also 
a {\it reverse} ratchet effect, in which 
the net dc vortex motion is in the hard substrate direction
\cite{Hastings,Hastings2,Zhu,Villegas,Triangular,Silva}.  
%NEW
%Symmetry breaking which requires 2D \cite{Hastings}.
%RCA is 1D \cite{Hastings2}.
%Analysis given in terms of a single particle; 1D effect \cite{Zhu}(a).
%A bottleneck effect \cite{Zhu}(b).
%A 1D picture \cite{Villegas}
%A truly 2D effect, could be called bottlenecking \cite{Triangular}.
%A 1D picture \cite{Silva}
%END NEW
In some cases, ratchet reversals occur when the vortex-vortex interactions 
at higher fillings become relevant, resulting in an effective
asymmetric potential oriented in the direction opposite to the pinning
structure.  As a result, some of the vortices  
move in the hard direction of the pinning substrate, and the net motion
is described by either an effectively 1D \cite{Zhu,Villegas,Silva} or 2D
\cite{Triangular} mechanism.
%\cite{Zhu,Villegas,Triangular,Silva}.  
Other vortex ratchet systems have multiple sign reversals 
due to symmetry breaking from multiple circular ac drives \cite{Hastings} 
or three state pinning potentials \cite{Hastings2}.  
The reverse vortex ratchet effect has been observed in 
experiments with 2D arrays of triangular pinning sites \cite{Villegas} 
and other types of 2D arrays \cite{Silva}. 
In the work of Silva {\it et al.} \cite{Silva}, the periodic pinning
array in the experimental sample has a two component unit cell containing
a large pinning site with a small pinning site placed to one side.
This system exhibits a remarkable 
field dependent {\it multiple} reverse ratchet effect.
For the first matching field and higher odd matching fields, 
a regular ratchet effect occurs, 
but at even matching fields the ratchet effect changes sign. 
Simulations indicate that the pinning sites are capturing
multiple vortices, and that this can change the direction of the 
asymmetric effective potential at every matching field, as
described by a 1D model. 
A ratchet effect has also been observed in  
a Josephson-junction array structure which has spatial
asymmetry \cite{Pastoriza}.  Here, a diode effect occurs 
where the depinning field is higher in the hard direction. 
A ratchet and reversed ratchet effect are also seen
for matching fields less than one.

\begin{figure}
\includegraphics[width=3.5in]{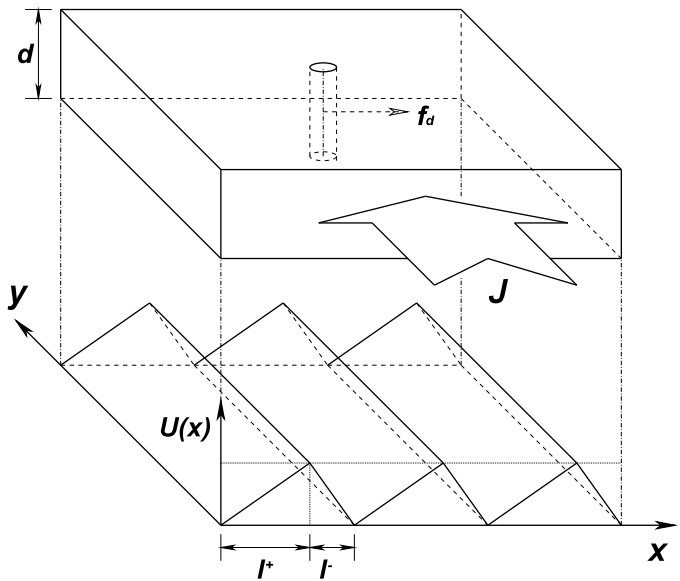}
\caption{
Schematic of the simulation geometry.  Vortices are subjected to an applied
current ${\bf J}=J{\bf {\hat y}}$ which produces a driving force
${\bf F}^{ext}=F_{dc}{\bf {\hat x}}$
or $F_{ac}{\bf {\hat x}}$, illustrated in the top portion of the figure.
The bottom portion of the figure shows the periodic 
%1D
potential $U(x)$ through which the vortices move.
The width of the easy direction of the pinning potential is $l^+$ and the
width of the hard direction is $l^-$, where $l^+>l^-$ and $l^++l^-=a$,
the pinning lattice constant.
}
\label{fig:schematic}
\end{figure}

In this work, we perform a detailed investigation of
the simplest vortex ratchet system created by an 
%1D
asymmetric periodic modulation, which was originally shown to
exhibit only an ordinary ratchet effect \cite{Janko}. 
By considering magnetic fields much higher than those used
in Ref.~\cite{Janko}, where the effectively single vortex regime
was studied, we show that this system can also exhibit ratchet sign reversals
as a function of vortex density when the vortex interactions become relevant. 
Unlike the majority of previously studied vortex ratchets, the behavior
of this system is determined by the detailed 2D structure of the vortex
configuration, and cannot be explained in terms of a simple 1D model.
We find that the dc response exhibits a reversible diode effect 
correlated with the regions where the ratchet effect reverses. 
The sign of the ratchet effect is shown to be related to the 
nature of the overall vortex lattice structure. 
At magnetic field densities where the vortex lattice has a more ordered 
or crystalline structure, the effective vortex interactions cancel or are
reduced by the symmetry of the lattice, and the response is the ordinary
ratchet effect as in the case of a single vortex.
At fields where the vortex lattice is disordered or has 
an intrinsic asymmetry, the interactions
between individual vortices become relevant 
and a portion of the vortices move in the 
%soft 
easy
direction of the pinning potential, resulting in a reversal of the sign
of the ratchet effect which can be predicted from force balance arguments.
In addition to the ratchet response of this system, we 
show that a rich variety of distinct vortex phases
can be realized as a function of density, including 
crystalline, smectic, disordered and square phases.
Commensuration effects also create oscillations in the critical 
depinning force which are distinct from those observed 
for two dimensional periodic pinning arrays. 

\begin{figure}
\includegraphics[width=3.5in]{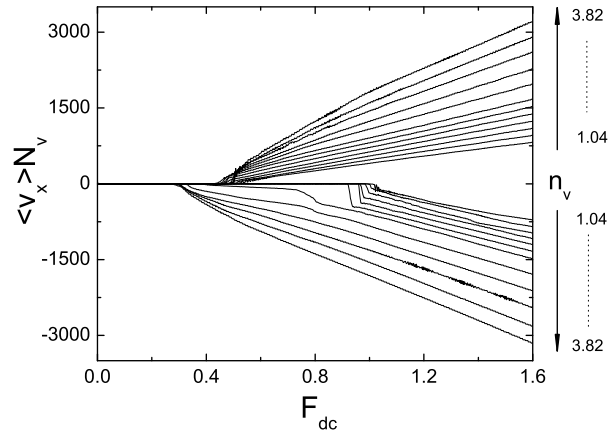}
\caption{
The dc non-normalized average velocity 
$\langle V_{x}\rangle N_v$ vs $F_{dc}$ for varied 
$n_{v}$ and both directions of dc drive. 
The curves for ${\bf F}_i^{ext}={\bf {\hat{x}}}F_{dc}$ 
have $\langle V_{x}\rangle N_v \ge 0$, while the curves for
${\bf F}_i^{ext}=-{\bf {\hat{x}}}F_{dc}$
have $\langle V_{x}\rangle N_v \le 0$.
In order of increasing $|\langle V_{x}\rangle N_v|$, the curves
have 
%$N_{v} =$ 600, 700, 800, 900, 
%1000, 1100, 1200,
%1400, 1600, 1800, 
%2000, and 2200. 
$n_v=1.04/\lambda^2$, $1.22/\lambda^2$, $1.39/\lambda^2$, $1.56/\lambda^2$,
$1.74/\lambda^2$, $1.91/\lambda^2$, $2.08/\lambda^2$,
$2.43/\lambda^2$, $2.78/\lambda^2$, $3.13/\lambda^2$,
$3.47/\lambda^2$, and $3.82/\lambda^2$.
For 
%$N_{v} < 1200$, 
$n_v<2.08/\lambda^2$, the negative drive critical depinning force 
is larger than the positive drive critical depinning force,
$f_c^->f_c^+$, but for 
%$N_v \geq 1200$, 
$n_v \geq 2.08/\lambda^2$, this reverses and
$f_c^-<f_c^+$.
}
\label{fig:dcdepin}
\end{figure}

\section{Simulation}
We consider a 2D system of size $L_x = L_y = 24\lambda$, where
$\lambda$ is the London penetration depth,
with periodic boundary conditions
in the $x$ and $y$ directions. 
The sample contains 
$N_v$ vortices interacting with
an 
%1D 
asymmetric substrate $U(x)$ of period $a=\lambda$,
illustrated schematically in Fig.~\ref{fig:schematic}.
The vortex density $n_v=N_v/L_xL_y$.
A given vortex $i$
obeys the overdamped equation of motion 
\begin{equation} 
{\eta} \frac{d {\bf R}_{i}}{dt} = {\bf F}^{vv}_{i} + {\bf F}^{s}_{i} +  {\bf F}^{ext}_{i} + {\bf F}^{T}_i .
\end{equation} 
The damping constant $\eta=\phi_0^2d/2\pi\xi^2\rho_N$, where $d$ is the
sample thickness, $\xi$ is the coherence length, $\rho_N$ is the normal-state
resistivity, and $\phi_0=h/2e$ is the elementary flux quantum.
The vortex-vortex interaction force is  
\begin{equation} 
{\bf F}^{vv}_{i} = \sum_{j \neq i}^{N_{v}}f_0 
K_1\left(\frac{r_{ij}}{\lambda}\right){\bf {\hat r}}_{ij}  
\end{equation} 
Here 
$r_{ij}=|{\bf r}_i-{\bf r}_j|$, 
${\bf {\hat r}}_{ij}=({\bf r}_i-{\bf r}_j)/r_{ij}$, and ${\bf r}_{i(j)}$ is
the position of vortex $i$ ($j$).
Force is measured in units of $f_{0} = \phi^{2}_{0}/2\pi\mu_{0}\lambda^{3}$,
%where $\phi_{0}$ is the elementary flux quantum. 
and time 
%is measured 
in units of $\tau = \eta/f_{0}$. 
$K_{1}(r_{ij}/\lambda)$ is the modified Bessel function 
which falls off exponentially for $r_{ij} > \lambda$.
For computational efficiency, the vortex-vortex interaction force is 
cut off at $6\lambda$. We have previously found that using
longer cutoff lengths produces negligible effects \cite{Bean}.   
The pinning potential is modeled as a sawtooth as 
shown in Fig.~\ref{fig:schematic}, 
\begin{displaymath}
{\bf F}_i^s=\left\{
\begin{array}{ll}
-\frac{1}{2}A_p{\bf {\hat x}} & \textrm{if $0 \le x_i\bmod a< l^+$}\\
A_p{\bf {\hat x}} & \textrm{if $l^+ \le x_i\bmod a < a$}.
\end{array} \right.
\end{displaymath}
Here, $A_p=1.0f_0$ and $x_i={\bf r}_i \cdot {\bf {\hat x}}$.
The width of the long or ``easy'' side of the pinning potential 
is $l^+=(2/3)a$ and the width of the short 
or ``hard'' side is $l^-=(1/3)a$ 
such that $l^++l^-=a$.  
The term ${\bf F}^{ext}_{i}$ in Eq.~1 is the force from a dc or ac
external driving current.
For the case of an applied dc drive, 
${\bf F}^{ext}_{i} = F_{dc}{\bf {\hat x}}$, where 
$F_{dc}$ is increased from zero in increments of 
$\delta F_{dc} = 0.01$ every $10^4$ simulation time steps.    
For the case of an applied ac drive,
${\bf F}^{ext}_{i}=\pm F_{ac}{\bf {\hat x}}$,
where the positive sign is used during the first half of each period $\tau$
and the negative sign is used during the second half period, resulting in
a square wave centered about zero.
The initial vortex configurations are obtained through simulated annealing,
where we set ${\bf F}_i^{ext}=0$ and apply a thermal force with the properties
$\langle {\bf F}_i^T\rangle=0$ and
$\langle {\bf F}_i^T(t){\bf F}_j^T(t^\prime)\rangle=
2\eta k_B T \delta_{ij} \delta(t-t^\prime)$.
The simulated annealing begins at a high temperature $F^T=5.0$ which is
well above the melting temperature of the vortex lattice, and then
$F^T$ is slowly
reduced to zero.
After annealing, we set ${\bf F}_i^T=0$, apply a dc or ac drive,
and measure the average velocity 
$\langle V_{x} \rangle=
\langle(N_v^{-1})\sum_{i=0}^{N_v}{\bf v}_i \cdot {\bf {\hat {x}}}\rangle$.
For dc driving, we identify the critical currents in the positive and
negative driving directions, $f_c^+$ and $f_c^-$, which we define as the
drive at which $|\langle V_x\rangle|=0.1f_0$. 

\begin{figure}
\includegraphics[width=3.5in]{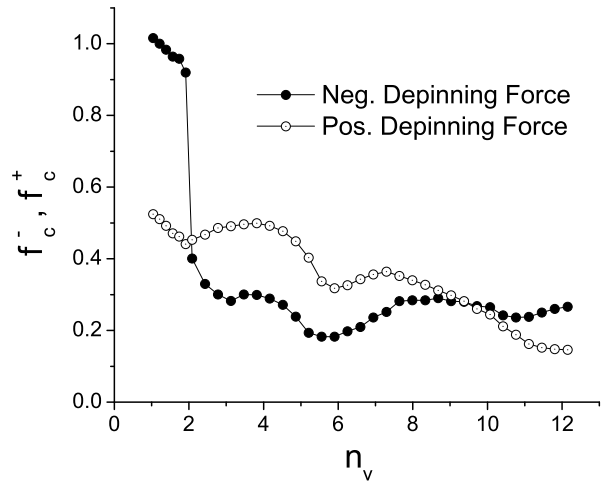}
\caption{
The depinning force curves vs 
%$N_{v}$  
$n_v$
for the system in Fig.~2 for driving in the negative
direction, $f_c^-$ (black circles), and in the positive direction, $f_c^+$ 
(open circles).  The
diode effect reverses near 
%$N_{v} = 1200$ and $N_{v} = 6000$.   
%$n_v=2.08/\lambda^2$ and $n_v=10.42/\lambda^2$.
$n_v=1.91/\lambda^2$ and $n_v=9.90/\lambda^2$.
Several oscillations in the depinning force also occur.
}
\label{fig:jc}
\end{figure}

\begin{figure}
\includegraphics[width=3.5in]{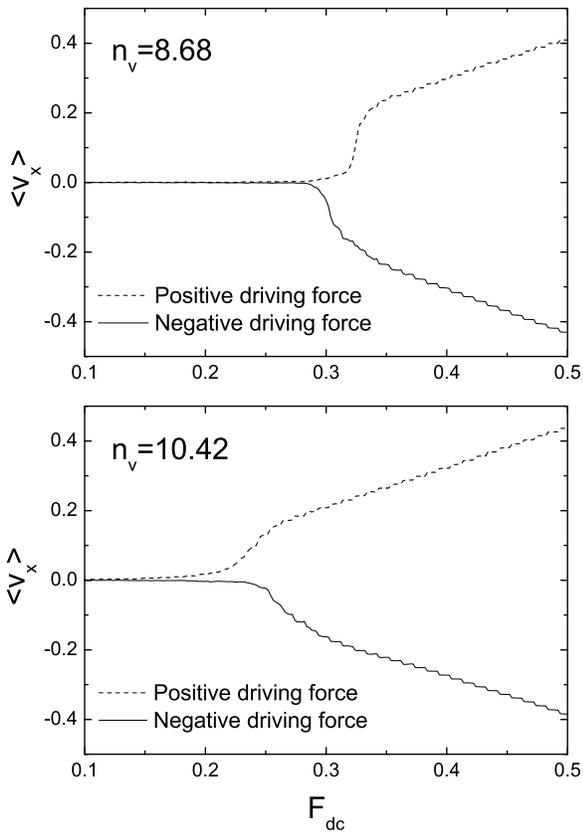}
\caption{  
$\langle V_x\rangle$ versus $F_{dc}$ curves for positive driving,
${\bf F}_i^{ext}={\bf {\hat x}}F_{dc}$ (upper dashed line),
and negative driving,
${\bf F}_i^{ext}=-{\bf {\hat x}}F_{dc}$ (lower solid line).
(a) A sample with 
%$N_v=5000$ 
$n_v=8.68/\lambda^2$
showing a negative diode effect with
$f_c^-<f_c^+$.
(b) A sample with 
%$N_v=6000$ 
$n_v=10.42/\lambda^2$
showing a positive diode effect with
$f_c^->f_c^+$.
}
\label{fig:iv}
\end{figure}

\begin{figure*}
\includegraphics[width=5.0in]{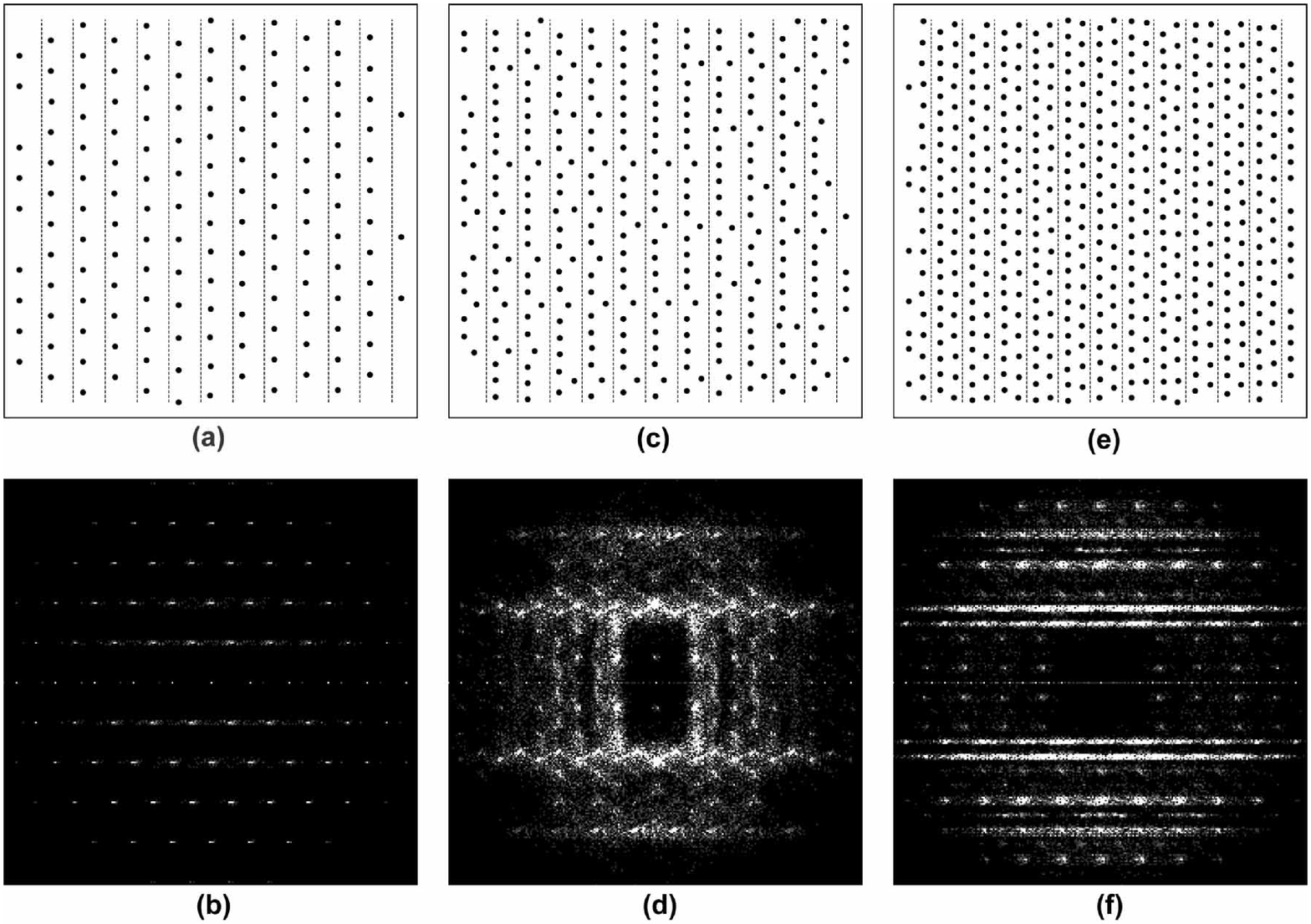}
\caption{  
(a) Vortex locations (black dots) and edges of the pinning troughs (black
lines) in a $12\lambda \times 12\lambda$ subsection of 
a system with 
%$N_v=600$ 
$n_v=1.04/\lambda^2$
and no applied drive.  A single
column of vortices is confined within each trough.
(b) Corresponding structure factor $S({\bf k})$ for 
%$N_v=600$.
$n_v=1.04/\lambda^2$.  Here and in other plots of $S({\bf k})$, we
show a greyscale heightmap of $S(k_x,k_y)$ with white values higher than
dark values, and with the origin
of coordinates at the center of the panel.  $k_x$ is along the vertical
axis, and $k_y$ is along the horizontal axis.
(c) Vortex and pinning trough locations for a system with 
%$N_v=1200$ 
$n_v=2.08/\lambda^2$
and
no applied drive.  In some regions, the columns of vortices have buckled and
a smectic type of ordering begins to appear.
(d) $S({\bf k})$ for 
%$N_v=1200$.
$n_v=2.08/\lambda^2$.
(e) Vortex and pinning trough locations for a system with 
%$N_v=2000$ 
$n_v=3.47/\lambda^2$
and no
applied drive.  Each trough contains two columns of vortices.
(f) $S({\bf k})$ for 
%$N_v=2000$.  
$n_v=3.47/\lambda^2$.
A double smectic ordering appears, as indicated
by the two prominent lines in $S({\bf k})$.
}
\label{fig:chain}
\end{figure*}

\section{DC Depinning and Vortex Lattice Structures}

We first compare the dc response of the system for driving
in the positive (easy) and negative (hard) $x$ direction.
The top set of curves in Fig.~\ref{fig:dcdepin} show
$\langle V_x\rangle N_v$ versus $F_{dc}$ for 
${\bf F}_i^{ext}={\bf {\hat x}}F_{dc}$ for samples with $N_v$ ranging
from 
%$N_v=600$ to 2200.  
$n_v=1.04/\lambda^2$ to $3.82/\lambda^2$.
The bottom set of curves show
$\langle V_x\rangle N_v$ versus $F_{dc}$ for the same samples with
${\bf F}_i^{ext}=-{\bf {\hat x}}F_{dc}$.
A clear diode effect occurs for 
%$N_v < 1200$ 
$n_v<2.08/\lambda^2$ 
when the negative (hard) critical
depinning force $f_c^-$ is larger than the positive (easy) critical
depinning force $f_c^+$.
Here, $f_c^-/f_0 \lesssim 1$, 
close to the value of the pinning force in the hard direction, while
$f_c^+/f_0 \approx 0.5$, nearly the same as the 
%soft 
easy direction pinning force.
Thus, for 
%$N_v < 1200$,  
$n_v<2.08/\lambda^2$,
the depinning force in each direction
is primarily determined by the 
pinning substrate and the vortex-vortex interactions are only weakly relevant. 
A pronounced drop in $f_c^-$ from $f_c^-/f_0 \approx 1$
to $f_c^-/f_0 \approx 0.4$ occurs above 
%$N_v=1200$.
$n_v=2.08/\lambda^2$.
For 
%$N_v > 1200$, 
$n_v>2.08/\lambda^2$,
the diode effect is reversed since now $f_c^-<f_c^+$.

When 
%$N_{v} = 1200$, 
$n_v=2.08/\lambda^2$, 
the negative depinning curve 
in Fig.~\ref{fig:dcdepin}
shows a clear two step depinning process with
an initial depinning near $F_{dc} = 0.45f_0$ followed by a second jump in
$\langle V_x\rangle$ near $F_{dc}=0.8f_0$.
At the low first depinning threshold, only a portion of the vortices depin,
while at the second threshold near $F_{dc}=0.8f_0$, the entire lattice depins.

%CIJOL Moving text
The values of $f_c^+$ and $f_c^-$ undergo 
changes as a function of 
%$N_{v}$ 
$n_v$ that are
difficult to discern in Fig.~\ref{fig:dcdepin}. 
In Fig.~\ref{fig:jc} we plot both
$f_c^+$ and $f_c^-$ versus 
%$N_v$ 
$n_v$
for systems with up to 7000 vortices and $n_v=12.15/\lambda^2$.
There is a positive diode effect for 
%$N_v<1100$ 
$n_v<1.91/\lambda^2$
when $f_c^->f_c^+$.
The diode effect reverses for
%$1100 \geq N_{v} \leq 5700$, 
$1.91/\lambda^2 \leq n_v \leq 9.90/\lambda^2$,
where $f_c^-<f_c^+$.
A second reversal occurs for 
%$N_v>5500$, 
$n_v>9.90/\lambda^2$,
when $f_c^+$ drops back below
$f_c^-$.
The high 
%$N_{v}$ 
$n_v$ reversal is shown in more detail in Fig.~\ref{fig:iv},
where we plot $\langle V_x\rangle$ versus $F_{dc}$ for two different samples.
In Fig.~\ref{fig:iv}(a), at 
%$N_v=5000$, 
$n_v=8.68/\lambda^2$,
$f_c^-<f_c^+$ and we find a
negative diode effect.  The situation is reversed in Fig.~\ref{fig:iv}(b) at
%$N_v=6000$, 
$n_v=10.42/\lambda^2$,
where $f_c^->f_c^+$ and a positive diode effect occurs.

\begin{figure*}
\includegraphics[width=5.0in]{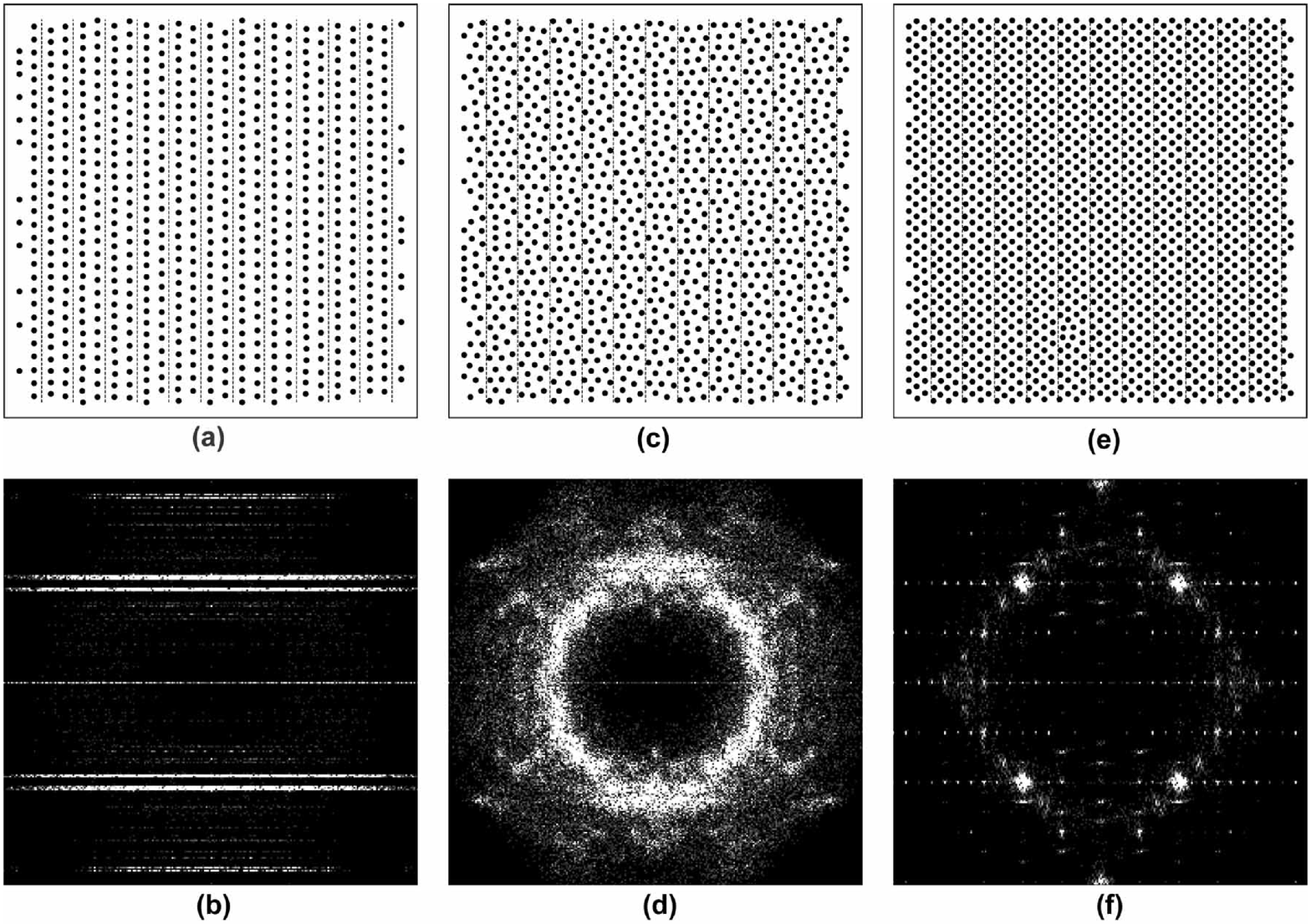}
\caption{ 
(a) Vortex locations (black dots) and edges of the pinning troughs 
(black lines) in a $12\lambda \times 12\lambda$ subsection of a system with
%$N_{v} = 3000$ 
$n_v=5.21/\lambda^2$
and no applied drive. Each trough captures two columns of
vortices.
(b) Corresponding $S({\bf k})$ for 
%$N_v=3000$ 
$n_v=5.21/\lambda^2$
showing 
evidence of smectic ordering. 
(c) Vortex and pinning trough locations for 
%$N_v=4000$.  
$n_v=6.94/\lambda^2$.
The vortex
configurations are disordered.
(d) $S({\bf k})$ for 
%$N_v=4000$ 
$n_v=6.94/\lambda^2$
has a clear ring structure.
(e) Vortex and pinning trough locations for 
%$N_v=6000$.  
$n_v=10.42/\lambda^2$.
The vortices
form a tilted square lattice.
(f) $S({\bf k})$ for 
%$N_v=6000$ 
$n_v=10.42/\lambda^2$ shows four-fold peaks.
}
\label{fig:smectic}
\end{figure*}

Fig.~\ref{fig:jc} indicates that $f_c^+$ and $f_c^-$ both undergo
oscillations.  
A simple 1D picture for the vortex structure would predict 
maxima to appear in the depinning force at the commensurate 
fields at which an integer number of vortex rows fit inside each pinning 
trough, given by $n_v=n^2/\lambda^2$, where $n$ is an integer. 
We do observe a maximum in both $f_c^+$ and $f_c^-$ near 
%CIJOL Changing to match reply.
%$N_{v} = 2600$ and $4500$. 
%$n_v=4.51/\lambda^2$ and $7.81/\lambda^2$.
$n_v=4.0/\lambda^2$,
% for both $f_c^+$ and $f_c^-$, where we expect to find
corresponding to the presence of 
two ordered commensurate vortex rows per channel.  
Although $f_c^-$ has a very broad plateau centered around
$n_v=9.0/\lambda^2$ which could be interpreted as the $n=3$ matching
field, the inadequacy of a simple 1D picture is indicated by the fact
that for $f_c^+$, the corresponding maximum falls at the much lower field
$n_v=7.81/\lambda^2$. 
%for $f_c^+$, while 
%These maxima result when the vortices form ordered commensurate structures 
%near 
%these densities.
For vortex densities just above and below each of 
the commensurate fields, the addition or subtraction of a small number
of vortices does not change the number of rows in each trough but instead
causes a disordering of the rows relative to one another, resulting in a
smectic structure.  This occurs because the shear modulus is smaller
than the compression modulus of the rows.  As larger numbers of vortices
are added, the compression of each row grows until the rows begin to buckle.
The buckling transition destroys the alignment perpendicular to the 
vortex rows, causing the vortex configuration to become more isotropically
disordered.
% and to develop a glasslike ordering.
This is similar to the commensuration effects found for vortex lattices 
in multilayer systems \cite{Layers}. 
In this case, the field is applied along the direction of the
layers so that the vortices effectively interact with a 
potential that is periodically modulated in one dimension. 
These experiments found oscillations in the magnetization, 
which is proportional to the depinning force.
The oscillations were also attributed to transitions in the 
vortex lattice structure from $n$ chains per layer to $n+1$ chains per
layer \cite{Layers}.  

The vortex configuration for 
%CIJOL This must be wrong
%$N < 1200$ 
%$n_v<2.08/\lambda^2$
$n_v<1.04/\lambda^2$
has only one column of vortices per 
pinning trough modulation, as illustrated in Fig.~\ref{fig:chain}(a).
The presence of triangular ordering in the vortex lattice is clearly
indicated by the structure factor $S({\bf k})$,
\begin{equation}
S({\bf k}) 
= N_{v}^{-1}\left|\sum^{N_{v}}_{i=1}\exp(i{\bf k}\cdot {\bf r}_{i})\right|^2 ,
\end{equation} 
plotted in Fig.~\ref{fig:chain}(b).
On average, the vortex-vortex interactions cancel along the $x$ direction since
the vortices are equally spaced a distance $a$ apart in this direction.
Near 
%$N=1200$, 
$n_v=2.0/\lambda^2$,
a transition in the vortex configuration occurs in which the
vortex columns begin to buckle, as shown in Fig.~\ref{fig:chain}(c).
Here, the vortex-vortex interactions along the $x$ direction no longer
cancel, and at points in the vortex columns where a buckled vortex is adjacent
to the column, a vortex within the column experiences both the force from the
pinning substrate, which has a maximum value of $A_p$, 
as well as an additional force from the buckled vortex, which has a maximum
value of $K_1(a)f_0$.  A simple estimate of the depinning 
force in a buckled region, $f_{c,b}^-$
gives $f_{c,b}^- = A_{p} - K_1(a)f_0 = 0.395f_0$, 
which is close to the value of $f_c^-$ shown in Fig.~\ref{fig:dcdepin} for
%$N>1200$.
$n_v>2.08/\lambda^2$.
For values of 
%$N_v$ 
$n_v$ just above the buckling onset,
buckling occurs in only a small number of places that are widely separated.
Vortices near the buckled locations depin at $f_{c,b}^-$, while vortices
outside of the buckled locations do not depin until $F_{dc}$  is close
to $A_p$.  The result is the two stage depinning process shown in 
Fig.~\ref{fig:dcdepin} for 
%$N_v=1200$.
$n_v=2.08/\lambda^2$.
As 
%$N_{v}$ 
the vortex density is further increased 
above 
%$N_v=1200$, 
$n_v=2.08/\lambda^2$, the distance between buckled regions decreases, smearing
out the two step depinning process.

For 
%$1100 < N_{v} < 3400$, 
$1.91/\lambda^2 < n_v < 5.90/\lambda^2$,
$y$ direction spatial correlations in the vortex
configurations are lost due to the buckling of the vortex columns, but
the pinning substrate maintains the spatial correlations in the $x$ direction,
resulting in smecticlike vortex structures.
This is illustrated in 
Fig.~\ref{fig:chain}(c,d) for 
%$N_v=1200$ 
$n_v=2.08/\lambda^2$
and in Fig.~\ref{fig:chain}(e,f) for
%$N_v=2000$.
$n_v=3.47/\lambda^2$.
At 
%$N_{v} = 1200$
$n_v=2.08/\lambda^2$
in Fig.~\ref{fig:chain}(c), a second column of vortices begins to form in each
pinning trough by means of a buckling transition, as previously discussed.
The corresponding structure factor in Fig.~\ref{fig:chain}(d)
shows some smectic type smearing, indicating that the
vortices are less correlated in the $y$-direction than in the $x$-direction.  
As $N_{v}$ increases, the correlations in the $y$ direction 
are further decreased, as seen in $S({\bf k})$ for 
%$N_v=2000$ 
$n_v=3.47/\lambda^2$
in 
Fig.~\ref{fig:chain}(f). 
At this density, the vortex configuration shown in Fig.~\ref{fig:chain}(e)
indicates that there are two columns of vortices per trough, 
which is also reflected by the double band feature in $S({\bf k})$. 
At 
%$N_{v} = 3000$, 
$n_v=5.21/\lambda^2$,
shown in Fig.~\ref{fig:smectic}(a), 
the vortex configuration still has two columns of vortices in each 
pinning trough, and the two band feature in $S({\bf k})$ is more prominent,
as illustrated in Fig.~\ref{fig:smectic}(b).
As $N_v$ increases further, the two column structure is destroyed and the
vortex configuration becomes highly disordered in both 
the $x$ and $y$ directions, as seen at 
%$N_v=4000$ 
$n_v=6.94/\lambda^2$
in Fig.~\ref{fig:smectic}(c).  Here $S({\bf k})$ has a ring structure, as shown
in Fig.~\ref{fig:smectic}(d).
The effect of the one-dimensional substrate can still
be seen in the form of small modulations along 
$k_{x}$ in Fig.~\ref{fig:smectic}(d). 

At 
%$N_{v} = 6000$, 
$n_v=10.42/\lambda^2$,
plotted in Fig.~\ref{fig:smectic}(e),
a new commensurate crystalline structure appears where four vortices 
fit across the pinning trough at a 45$^\circ$ angle so that the
overall vortex lattice symmetry is square.
This is supported by the four prominent peaks that appear in $S({\bf k})$ in
Fig.~\ref{fig:smectic}(f).
%NEW
There is a competition between minimizing
the vortex-vortex interaction energy, which favors a triangular lattice,
and maximizing the vortex-substrate interaction, which favors commensurate
vortex configurations.  Since the difference in energy between a triangular
and square vortex lattice is quite small, the commensuration effect is
favored at $n_v=10.42/\lambda^2$, and a square structure appears.
%END NEW
%The vortex lattice configuration has a square symmetry at 
%%$N_v=6000$ 
%$n_v=10.42/\lambda^2$
%as shown
%in Fig.~\ref{fig:smectic}(e), producing four 
%prominent peaks in the structure factor
%$S(k)$ which is illustrated in Fig.~\ref{fig:smectic}(f).
Since the vortex lattice is symmetrical, the vortex-vortex interactions 
are strongly reduced and the depinning transition is
similar to the case of 
%$N_{v} < 1100$ 
$n_v<1.91/\lambda^2$
where $f_c^->f_c^+$ and the vortices
depin more readily in the 
%soft 
easy direction.
In this case, the depinning response is elastic, but since there are multiple
vortex columns in each trough there is a pronounced distortion in the lattice
when a dc force is applied, causing $f_c^-$ to be 
significantly lower than $A_p$.

\begin{figure}
\includegraphics[width=3.5in]{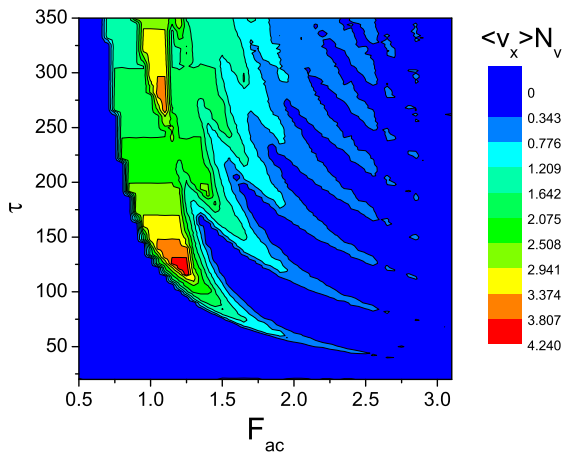}
\caption{(Color online)
Contour plot of $\langle V_{x}\rangle$ as a function of ac drive period
$\tau$ (in units of simulation time steps) and ac drive amplitude $F_{ac}$ 
(in units of $f_0$) for
a system with 
%$N_{v} = 50$.    
$n_v=0.087/\lambda^2$.
}
\label{fig:contour}
\end{figure}

As a function of vortex density, the system thus passes through a 
sequence of triangular, smectic, disordered, and square vortex arrangements.
These results show similarities to studies 
performed on magnetic particles interacting 
with simple symmetric periodic 
%1D 
modulated substrates, 
where crystalline, smectic, and partially disordered
structures were observed as a function of particle density \cite{Westervelt}. 
The double smectic phases and the square lattice that we find did not
appear in the magnetic particle experiment since the relative 
particle densities considered in Ref.~\cite{Westervelt} were much
lower than the relative vortex densities we use here. 
Additionally, the asymmetry in our potential substrate 
may help to stabilize certain phases that are not seen with 
simple symmetric potentials.   

\section{Ratchet and Reverse Ratchet Effect} 

We next examine the response when an external ac drive is applied. 
When the vortex density is low, the vortex-vortex interactions 
are negligible and the system responds in the single particle limit. 
In Fig.~\ref{fig:contour} we show a contour plot of the
average velocity $\langle V_{x}\rangle$ 
obtained from simulations with 
%$N_v=50$ 
$n_v=0.087/\lambda^2$
for varied ac period
$\tau$ and ac amplitude $F_{ac}$. 
A series of tongues where a positive ratchet effect occurs are
clearly visible.
At small $\tau$, there is not enough time for the vortices to respond to the
ac drive so there is no ratchet effect. Similarly, at 
low $F_{ac}$, the vortices are unable to cross the potential
barrier of the trough, and no rectification occurs. 
At intermediate values of $\tau$ and $F_{ac}$, a ratchet effect 
can occur.  
In the first tongue, which falls at the lower left edge of
Fig.~\ref{fig:contour}, the vortices move by $+a$ in the $x$ direction
during the positive half of the driving period, but do not move back
during the negative half of the period, resulting in a net rectification.
On the second tongue, vortices translate by $+2a$ during the first half
of the period and by $-a$ during the second half of the period, again
giving a net rectification.  
As $\tau$ and $F_{ac}$ increase,
the ratchet effect on the $n$th tongue is
characterized by motion of $+na$ during the positive half period of 
the drive and motion of $-(n-1)a$ during the negative half period of the
drive.
The tongue structure emerges since for some values of $\tau$ and
$F_{ac}$, vortices undergo equal translations during both halves of the
driving period, leading to no net motion.
The boundary of the $n$th tongue where ratcheting can occur is given by
\begin{equation}
\tau_{n} = \eta \frac{nl^+}{F_{ac} - F^+} + \eta 
\frac{(n - 1)l^-}{F_{ac} + F^-}    
\end{equation} 
The weak pinning force $F^+$ associated with the trough side of 
length $l^+$ is $F^+=-A_p/2$ in our case, while the strong pinning
force $F^-$ associated with the trough side of length $l^-$ is
$F^-=A_p$.
For the first tongue in Fig.~\ref{fig:contour}, Eq. 4 predicts
$\tau_{1} \propto (F_{ac} + A_p/2)^{-1}$, which matches the simulation results.
The results in Fig.~\ref{fig:contour} and Eq.~4 agree well
with the tonguelike features found in Ref.~\cite{Janko} as a function
of $\tau$ and $F_{ac}$.
This confirms that at low vortex densities, an
ordinary ratchet effect occurs where the vortices move in the easy direction. 

\begin{figure}
\includegraphics[width=3.5in]{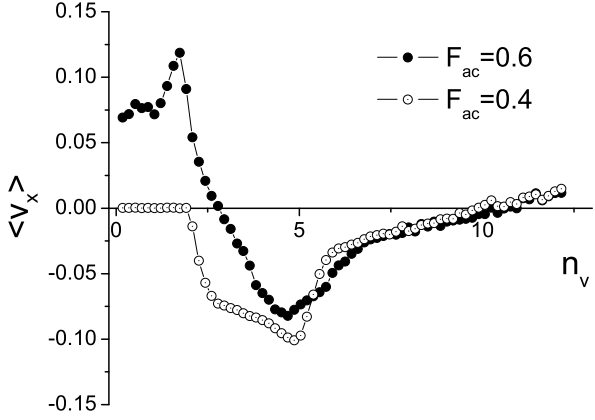}
\caption{  
$\langle V_{x}\rangle$ 
vs 
%$N_{v}$ 
$n_v$ for the system in 
Fig.~\ref{fig:contour} at fixed $\tau=10^4$ simulation steps
for (black circles) $F_{ac} = 0.6f_0$ and
(open circles) $F_{ac} = 0.4f_0$.
Here both positive
and negative ratchet effects occur as a function of vortex density.    
}
\label{fig:vxnv}
\end{figure}

\begin{figure}
\includegraphics[width=3.5in]{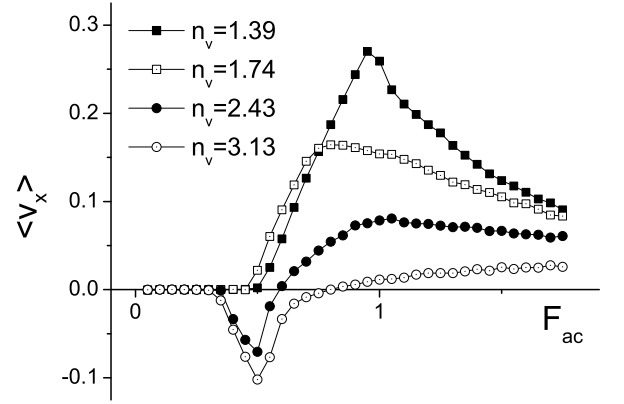}
\caption{$\langle V_{x}\rangle$ vs $F_{ac}$ 
for the same system as in Fig.~\ref{fig:vxnv} for varied 
%$N_{v}$. 
$n_v$.
Filled squares: 
%$N_v=800$; 
$n_v=1.39/\lambda^2$;
open squares: 
%$N_v=1000$; 
$n_v=1.74/\lambda^2$;
filled circles:
%$N_v=1400$; 
$n_v=2.43/\lambda^2$;
open circles: 
%$N_v=1800$.
$n_v=3.13/\lambda^2$.
The positive ratchet effect is preceded by a negative
ratchet effect for 
%$N_{v} > 1000$. 
$n_v>1.74/\lambda^2$.
}
\label{fig:vxac}
\end{figure}

We next consider the effect of increasing 
%$N_{v}$. 
$n_v$.
Since the ratchet effect in Fig.~\ref{fig:contour} is enhanced for larger
periods $\tau$, we fix the period at $\tau=10^4$ simulation steps.
In Fig.~8 we plot $\langle V_{x}\rangle$ versus  
%$N_{v}$
$n_v$
for $F_{ac} = 0.6f_0$ and $0.4f_0$. For 
%$N_{v} < 1500$, 
$n_v<2.60/\lambda^2$,
there is a 
positive ratchet effect
for $F_{ac} = 0.6f_0$ and no ratchet effect for $F_{ac} = 0.4f_0$.
For 
%$N_{v} > 1500$ 
$n_v>2.60/\lambda^2$
a negative ratchet effect
occurs for $F_{ac} = 0.4f_0$, while the 
positive ratchet effect for $F_{ac} = 0.6f_0$ decreases in size and reverses
to a negative ratchet effect near 
%$N_{v} = 2100$.
$n_v=3.65/\lambda^2$.
For both values of $F_{ac}$, the system passes through a maximum
negative value of $\langle V_x\rangle$ just below 
%$N_{v} = 3000$. 
$n_v=5.21/\lambda^2$.
Above 
%$N_v=3000$, 
$n_v=5.21/\lambda^2$,
$\langle V_x\rangle$ gradually increases back toward
zero before crossing zero and reversing to a small positive ratchet effect
near 
%$N_{v}=6000$.
$n_v=10.42/\lambda^2$.
The features in these curves appear to be correlated with the vortex lattice structures presented in Figs~\ref{fig:chain} and \ref{fig:smectic}.
At low 
%$N_{v}$, 
$n_v$,
the vortices form a triangular or smectic structure 
with only a single column of vortices in each pinning trough, as in
Fig.~\ref{fig:chain}(a,b).
In this case, since the symmetry of the 
vortex configuration cancels or strongly reduces the 
vortex-vortex interaction force, a positive ratchet effect occurs 
that is determined only by the pinning forces of the substrate.  
The onset of the negative ratchet effect corresponds to the appearance of
buckling of the columns of vortices in the pinning troughs, as seen in
Fig.~\ref{fig:chain}(c,d).
In the double smectic phase illustrated in Fig.~\ref{fig:chain}(e,f) 
and Fig.~\ref{fig:smectic}(a,b), the negative ratchet
effect persists.
The gradual disappearance of the negative ratchet effect 
for 
%$N_{v} > 3000$ 
$n_v>5.21/\lambda^2$
in Fig.~\ref{fig:vxnv} is associated 
with the crossover from the double smectic state to the 
disordered state shown in Fig.~\ref{fig:smectic}(c,d). The
small positive ratchet effect near 
%$N_{v} = 6000$ 
$n_v=10.42/\lambda^2$
is correlated with the formation of the ordered square 
lattice as seen in Fig.~\ref{fig:smectic}(e,f). 

In Fig.~\ref{fig:vxac} we plot $\langle V_{x}\rangle$ versus $F_{ac}$ 
for a series of simulations performed at varied
%$N_{v}$. 
$n_v$.
For low 
%$N_{v} \le 1000$, 
$n_v \le 1.74/\lambda^2$,
there is no ratchet effect until a 
threshold value of $F_{ac}$ is reached.
$\langle V_{x}\rangle$ then increases to a peak value and gradually
decreases with increasing $F_{ac}$.
For 
%$N_{v} > 1000$, 
$n_v>1.74/\lambda^2$,
the threshold value of $F_{ac}$ above which a ratchet
effect occurs is lower than for 
%$N_v \le 1000$, 
$n_v \le 1.74/\lambda^2$,
and the initial ratchet
effect is in the negative direction rather than the positive direction.
As $F_{ac}$ is increased, the negative ratchet effect 
reverses to a positive ratchet effect.     

\begin{figure}
\includegraphics[width=3.5in]{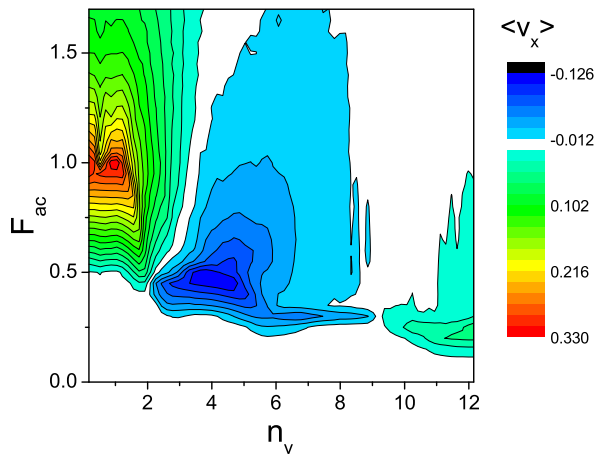}
\caption{(Color online)  
A colorscale map of $\langle V_x\rangle$ as a function of
$F_{ac}$ and 
%$N_v$.
$n_v$.
For low $F_{ac}$ there is no ratchet effect. 
At higher $F_{ac}$ there is a positive ratchet effect for
low 
%$N_{v}$, 
$n_v$,
a reversal to a negative ratchet effect for intermediate 
%$N_{v}$, 
$n_v$,
and another reversal at high 
%$N_{v}$ 
$n_v$
to a 
positive ratchet effect.   
}
\label{fig:map}
\end{figure}

Figure \ref{fig:map} 
illustrates $\langle V_x\rangle$ in a contour plot as a function
of $F_{ac}$ and 
%$N_v$.
$n_v$.
The positive and negative ratchet effect regimes are highlighted. 
For low $F_{ac}$ there is no ratchet effect, while for 
high $F_{ac}/f_0 > 1.0$ 
all ratchet effects are gradually reduced.  The maximum 
positive ratchet effect occurs for $F_{ac}/f_0$ close to $1.0$, 
the value of the pinning
force in the strong direction. 
The positive ratchet effect
for 
%$N_{v} < 1500$ 
$n_v<2.60/\lambda^2$
does not occur unless 
$F_{ac}/f_0>0.5$, which is the maximum
pinning force in the weak direction. 
The negative ratchet effect appears for 
%$N_{v} > 1200$ 
$n_v>2.08/\lambda^2$
and
extends up to 
%$N_{v} = 5000$, 
$n_v=8.68/\lambda^2$,
with the maximum negative rectification occurring for 
%$2000 < N_{v} < 3000$ 
$3.47/\lambda^2 < n_v < 5.21/\lambda^2$
at $0.45 < F_{ac}/f_0 < 0.5$.
For 
%$N_{v} > 5600$, 
$n_v>9.72/\lambda^2$,
a reentrant positive ratchet effect appears that is 
much weaker in 
amplitude than the positive ratchet effect region found for low 
%$N_v$. 
$n_v$.
In general, the ratchet effect in either direction 
is reduced for increasing 
%$N_{v}$ 
$n_v$ since the
effects of the substrate are gradually washed out by the
increasing vortex-vortex interaction forces. 
This result shows that vortices interacting with  simple asymmetric pinning 
potentials can exhibit multiple ratchet reversals
similar to those seen for the more complex 2D pinning array geometries 
\cite{Silva}. It is likely that for
higher 
%$N_{v}$,
$n_v$, higher order reversals will occur
when the vortices again form a smectic phase with five or more vortex 
chains per pinning trough.  We are not able to access the high vortex
densities where these reversals would be expected to appear.
The higher order ratchet effects would
likely continue to decrease in amplitude with increasing field.   

\begin{figure}
\includegraphics[width=3.5in]{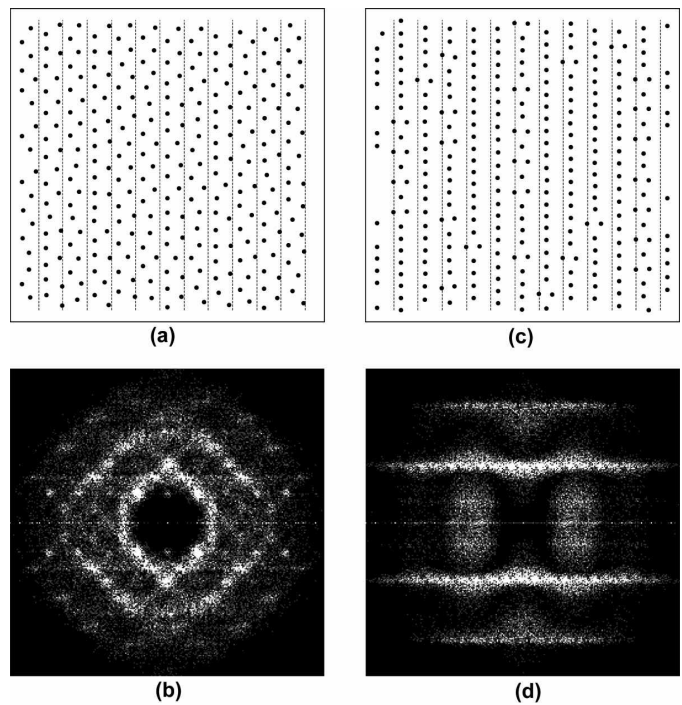}
\caption{  
(a) Vortex locations (black dots) and edges of the pinning troughs (black
lines) in a $12\lambda \times 12\lambda$ subsection of a system with
%$N_{v} = 1200$ 
$n_v=2.08/\lambda^2$
at $F_{ac}=0.5f_0$ when the negative ratchet effect occurs.
The positive portion of the driving period, with 
${\bf F}_i^{ext}={\bf {\hat x}}F_{ac}$, is shown.
The vortices form two columns in every trough.
(b) The corresponding structure factor $S({\bf k})$ for positive drive.  The 
overall disorder in the vortex configuration is indicated by the ring
structure. 
(c) Vortex and pinning trough locations for the same system during the
negative portion of the driving period, with
${\bf F}_i^{ext}=-{\bf {\hat x}}F_{ac}$.
The vortices are arranged in buckled columns.
(d) The corresponding $S({\bf k})$ for negative drive shows a smectic type of
ordering.
}
\label{fig:negative}
\end{figure}

The positive or regular ratchet effect that occurs 
when the vortices move in the easy direction is straightforward to 
understand at low fields based on a single particle picture.  
The first reversal of the ratchet effect for increasing $N_v$ 
is related to the vortex-vortex interactions.
This is more clearly seen from Fig.~\ref{fig:negative}, 
where we plot the vortex positions and corresponding structure factor
during both halves of the driving cycle
for 
%$N_{v} = 1200$ 
$n_v=2.08/\lambda^2$
and $F_{ac}=0.5f_0$ 
where a negative ratchet effect occurs.   
During the positive drive portion of the ac cycle, the vortices are
roughly evenly spaced in a disordered arrangement, as shown in
Fig.~\ref{fig:negative}(a,b).
When the negative drive is applied, the vortices form a clear smectic structure
with roughly one vortex column per pinning trough, as
illustrated in Fig.~\ref{fig:negative}(c,d).
In this case, the vortex configuration is asymmetric due to the appearance
of buckling in the vortex columns.
At a buckled location, an extra vortex appears on the positive $x$ side 
of the vortex column (to the right of the column) in the same trough.
These extra vortices effectively push the adjacent vortices in the column
over the potential barrier in the hard or negative $x$ direction. 

If the extra vortex is located a distance $a$ from the column,
then a portion of the vortices in the column can escape over the potential
barrier in the hard direction when the following condition is met: 
\begin{equation}
F_{ac} +K_1(a)f_0 > A_{p}. 
\end{equation}
When $A_{p}/f_0 = 1.0$ and $K_{1}(a) = 0.6$, 
a negative ratchet effect is predicted to occur when $F_{ac}/f_0 > 0.4$,
which is what is observed in Fig.~\ref{fig:vxac}. 
It could also be argued that when the drive is in the positive direction,
the extra vortices should be pushed by the column of vortices and should
more easily move in the positive direction, allowing
the conditions of Eq.~5 to be met in the positive direction as well. 
In Fig.~\ref{fig:negative}(a) we show that this does not occur because the
vortex arrangement is quite different during the positive drive portion of
the cycle.  The vortex-vortex interaction term is not present since the 
vortices do not form the smectic structure that is seen when the
driving is in the negative direction. Instead, the vortices 
form two columns within each pinning trough, allowing the individual 
vortices to remain evenly separated in both the $x$ and $y$ directions, and
preventing an individual vortex from pushing another vortex out of the
trough.
The vortex structure for positive drive is more symmetrical, 
as seen in $S({\bf k})$ in Fig.~\ref{fig:negative}(b) which
has a smeared crystalline or disordered structure. 
This symmetry reduces the vortex-vortex interaction forces.

A ratchet effect in the positive direction occurs 
when the following equation is satisfied: 
\begin{equation}
F_{ac} > A_{p}/2 .
\end{equation}
If $F_{ac}/f_0 = 0.45$ and $A_{p}/f_0 = 1.0$, then the conditions of 
Eq.~5 are met but the conditions of
Eq.~6 are not, so the negative ratchet effect 
can occur but not the positive ratchet effect.   
Once $F_{ac}/f_0 > 0.5$, it is also possible for the positive ratchet effect
to occur and the two ratchet mechanisms will compete, decreasing the
magnitude of the negative ratchet effect as seen in the 
phase diagram of Fig.~\ref{fig:map} for higher $F_{ac}$. 
As 
%$N_{v}$ 
$n_v$ is increased, the
smectic structure is gradually lost as shown in Fig.~\ref{fig:smectic}(c,d), 
so the vortex-vortex interactions required to produce the negative ratchet
effect are reduced and the negative ratchet effect is diminished.
Once the vortex lattice is again symmetrical, 
as in Fig.~\ref{fig:smectic}(e,f), the positive ratchet effect appears.
This result suggests that in regions where a reversed ratchet effect 
appears, the vortex lattice structure must exhibit some asymmetry 
when driven in the hard direction for the reverse ratchet effect to occur.     

\begin{figure}
\includegraphics[width=3.5in]{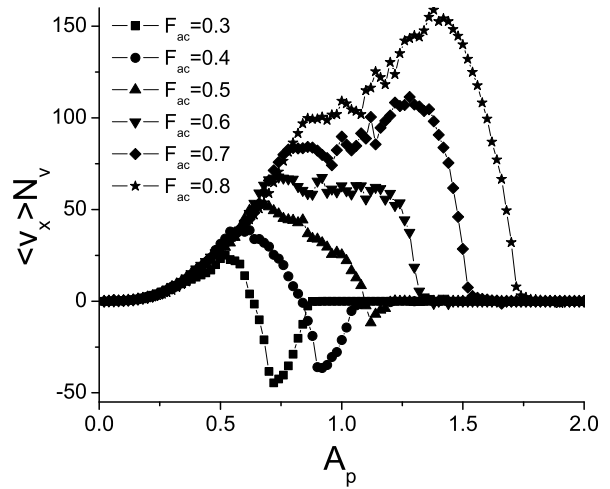}
\caption{  
$\langle V_{x}\rangle N_v$ vs $A_{p}$ 
for the same system as in Fig.~\ref{fig:map} with 
%$N_{v} = 1800$.
$n_v=3.13/\lambda^2$.
Squares: $F_{ac}=0.3f_0$; circles: $F_{ac}=0.4f_0$;
up triangles: $F_{ac}=0.5f_0$; down triangles: $F_{ac}=0.6f_0$;
diamonds: $F_{ac}=0.7f_0$; stars: $F_{ac}=0.8f_0$.
For large $A_{p}$ only a positive vortex ratchet effect occurs 
as the vortices form a single column in each pinning trough. 
As the pinning strength is reduced, buckling of the columns
begins to occur and a negative ratchet effect appears.
}
\label{fig:buckle}
\end{figure}

We have also investigated the reversal of the ratchet effect as a 
function of the pinning strength $A_{p}$ for fixed 
%$N_{v} = 1800$ 
$n_v=3.13/\lambda^2$
and $\tau = 10^4$.
In Fig.~\ref{fig:buckle} we plot $\langle V_{x}\rangle N_v$ 
versus $A_p$ for $F_{ac}$ ranging from $0.4f_0$ to $0.8f_0$.
For all values of $F_{ac}$, as $A_{p}$ increases a positive ratchet
effect appears.
This is because the stronger $A_{p}$ forces
the vortices to form completely 1D unbuckled columns inside the pinning
troughs.
For lower $A_{p}$, the vortex columns can buckle, providing the
unbalanced vortex-vortex interactions required to produce the negative
ratchet effect. 
The negative rectification is enhanced for both low $A_{p}$ and low $F_{ac}$.
Changing the strength of $A_{p}$ in a single superconducting sample
is difficult;  however, a geometry similar to that shown in 
Fig.~\ref{fig:schematic} could be created using tunable optical traps 
for colloidal particles driven with an ac electric field. 

\section{Summary} 

In conclusion, we have shown that the vortex ratchet system proposed by 
Lee {\it et al.} in Ref.~\cite{Janko} for
vortices interacting with simple periodic asymmetric pinning substrates
can exhibit a series of ratchet reversals similar to those observed
for 2D asymmetric periodic pinning arrays. 
Unlike most vortex ratchet geometries considered to date, the ratchet
effects presented here can only be explained by a 2D description of the
vortex configuration, and not by a simple 1D model.
As a function of vortex density, a rich variety of vortex phases 
can be realized in this system including triangular, smectic, disordered,
and square lattices. 
The dc critical depinning forces exhibit a reversible diode effect 
as a function of vortex
density.  At low magnetic fields, the depinning force is higher in the 
hard direction, corresponding to the regular diode effect, 
while at higher magnetic fields, the depinning force is 
lower in the hard direction.  Another reversal of the diode effect
occurs at even higher fields. 
The reversed diode effect appears when the symmetry of the vortex
configuration changes for different directions of dc drive, as in the
case of the smectic and disordered phases.  Here, the importance of
vortex-vortex interactions differs for the two drive directions.
At densities where the vortex lattice is highly ordered, a regular diode 
effect occurs where the vortex depinning force is lower  
in the easy direction. 
The vortex lattice structures are also directly correlated with the sign 
of the ratchet effect that occurs under an ac drive.  The highly symmetric
structures ratchet in the easy direction since 
the vortex-vortex interactions effectively cancel, causing the
sign of the ratchet effect to be determined only by the 
asymmetry of the pinning substrate. In the smectic states,
the vortex-vortex interactions become relevant and can produce a 
negative ratchet effect, which can 
be predicted with force balance arguments.
We have shown the dependence of the reversible ratchet effect on 
ac amplitude, vortex density, and pinning strength. 
This type of ratchet could be also realized for other kinds of systems 
composed of collections of repulsively interacting particles and 
simple asymmetric substrates, such as colloid systems.      

\section{Acknowledgments}
This work was carried out under the auspices of the National Nuclear
Security Administration of the U.S. Department of Energy at Los Alamos
National Laboratory under Contract No.~DE-AC52-06NA25396.  Q.L. was
also supported in part by NSF Grant No.~DMR-0426488.

\end{document}